\documentclass[aps,prl,superscriptaddress,showpacs,twocolumn]{revtex4-1}

\usepackage{graphicx}
\usepackage{amsmath}
\usepackage{dsfont}
\usepackage{xspace}
\usepackage[usenames,dvipsnames]{color}
\usepackage[colorlinks=true,linkcolor=blue,urlcolor=blue,citecolor=blue]{hyperref}

\newcommand{\ie}{i.e.\@\xspace}
\newcommand{\eg}{e.g.\@\xspace}

\newcommand{\eqw}[1]{(\ref{#1})}
\newcommand{\eq}[1]{Eq.\thinspace{}(\ref{#1})}
\newcommand{\eqq}[2]{Eqs.\thinspace{}(\ref{#1}) and (\ref{#2})}

\newcommand{\fig}[1]{Fig.\thinspace{}\ref{#1}}

\newcommand{\fc}[1]{({#1})}
\newcommand{\figc}[2]{Fig.\thinspace{}\ref{#1}\thinspace{}\fc{#2}}

\begin{document}

\title{Transport in two-dimensional disordered semimetals}

\author{Michael Knap}
\email[]{knap@physics.harvard.edu}
\affiliation{Department of Physics, Harvard University, Cambridge MA 02138, USA}
\affiliation{ITAMP, Harvard-Smithsonian Center for Astrophysics, Cambridge, MA 02138, USA}

\author{Jay D. Sau}
\affiliation{Department of Physics, Harvard University, Cambridge MA 02138, USA}
\affiliation{Department of Physics, Condensed Matter Theory Center and the Joint Quantum Institute, University of Maryland, College Park, MD 20742, USA}

\author{Bertrand I. Halperin}
\affiliation{Department of Physics, Harvard University, Cambridge MA 02138, USA}

\author{Eugene Demler}
\affiliation{Department of Physics, Harvard University, Cambridge MA 02138, USA}

\date{\today}

\begin{abstract}

We theoretically study transport in two-dimensional semimetals. 
Typically, electron and hole puddles emerge in the transport layer of these systems
due to smooth fluctuations in the potential. We calculate the electric response of 
the electron-hole liquid subject to zero and finite perpendicular magnetic fields 
using an effective medium approximation and a complimentary mapping on resistor networks. 
In the presence of smooth disorder and in the limit of weak electron-hole recombination 
rate, we find for small but finite overlap of the electron and hole bands an abrupt 
upturn in resistivity when lowering the temperature but no divergence at zero temperature. 
We discuss how this behavior is relevant for several experimental realizations and 
introduce a simple physical explanation for this effect.

\end{abstract}


\maketitle

In semimetals both electrons and holes contribute to transport. Typical examples
are indirect bulk semiconductors with small band overlap. More recently
also two-dimensional systems, including HgTe quantum wells close to the topological 
insulator to metal transition~\cite{kvon_two-dimensional_2008,olshanetsky_scattering_2009,
kvon_two-dimensional_2011,olshanetsky_two-dimensional_2012,olshanetsky_metal_2013}, 
BiSe thin films~\cite{kastl_local_2012}, and bilayer graphene~\cite{feldman_broken-symmetry_2009,
mccann_electronic_2013} have been identified to exhibit semimetallic properties.
Electron and hole puddles typically emerge in the 
transport layer of these systems due to disorder that varies smoothly in space on a 
scale that is large compared to the mean free path of the charge carriers.

\begin{figure} 
\begin{center}
 \includegraphics[width=0.48\textwidth]{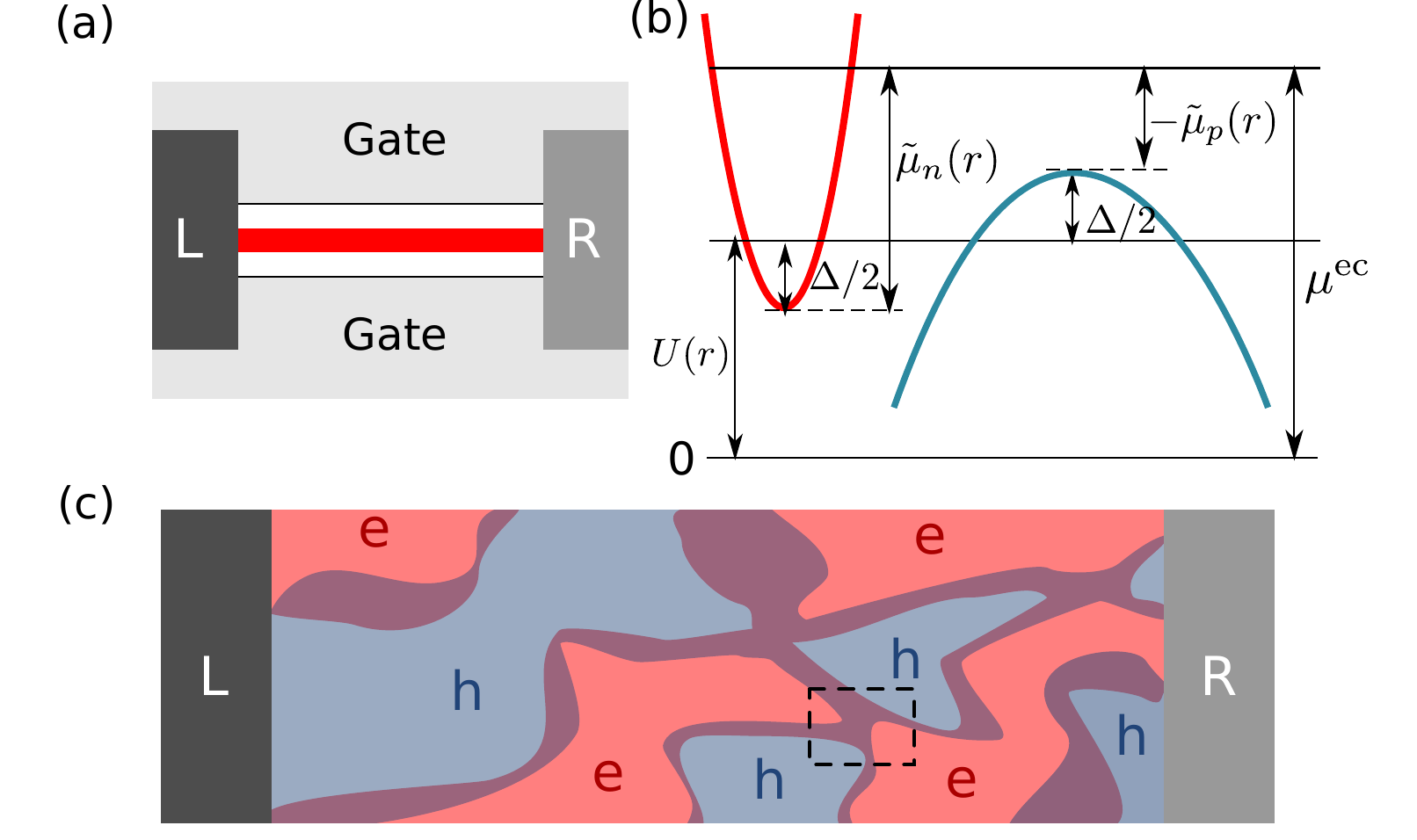}
\end{center}
\caption{\label{fig:setting} (Color online) \fc{a} A two-dimensional electron and hole mixture, red region, is driven
by a bias voltage imprinted from the potential difference in the left and the right lead [side view]. 
\fc{b} Definitions of the energy scales: 
$\Delta$ is the distance between  the conduction and the valance band edge which is taken to be negative when they overlap.
$\mu^\text{ec}$ is the electrochemical potential which at equilibrium is identical for electrons and holes. 
The electrical potential $U(r)$ and thus also the chemical potentials of the electrons $\tilde \mu_n(r)$ and holes 
$\tilde \mu_p(r)$ vary smoothly in space due to the disorder. \fc{c} The long-ranged fluctuations
in the potential create regions where only electrons (red) or only holes (blue) are occupied, and regions 
where both carriers coexist (purple) [top view].
}
\end{figure}

In this Letter, we introduce a two-fluid model to explore the effects of smooth 
disorder on the transport properties of two-dimensional
electron-hole mixtures, \fig{fig:setting}. 
The model assumes that as a result of long-range correlated
disorder, and a small intrinsic overlap between the electron and hole
band, the sample at low temperatures can be divided into three types
of regions: areas where only electron states or only hole states are
occupied, and intermediate areas where both types of carriers are
occupied, \figc{fig:setting}{c}. We assume that carriers are scattered easily within a
band, due to phonons or residual impurities, but that recombination
between electrons and holes is suppressed.  Then, if neither the pure
electron regions or pure hole regions percolate across the sample, charge
transport at low temperatures may be effectively limited by the relatively
narrow percolating portion where both electrons and holes are occupied
and each carrier type has low density. 
As we shall see, this can lead to an anomalously high resistivity at low
temperatures.

To make our picture quantitative, we introduce a model with smooth disorder, 
and obtain the local densities of
electrons and holes using a Thomas-Fermi-like  approximation.  We assume
that the electron and hole mobilities differ from each other, but are
independent of the respective carrier densities. We solve the
resulting inhomogeneous, two-component conductance problem using an effective medium
approximation (EMA)~\cite{bruggeman_berechnung_1935,landauer_electrical_1952,
stroud_generalized_1975,kirkpatrick_percolation_1973,rossi_effective_2009,tiwari_model_2009}. EMA has already been used successfully to
characterize transport in GaAs quantum wells, where smooth disorder has
been identified as the main mechanism for the transition from metallic to
insulating behavior as a function of electron density~\cite{kravchenko_possible_1994,
meir_percolation-type_1999,ilani_unexpected_2000,
meir_two-species_2000,meir_zero-field_2001,das_sarma_two-dimensional_2005,
tracy_observation_2009,das_sarma_two-dimensional_2013}. We investigate dependences of the transport on
band-overlap, temperature, gate voltage, electron-hole
recombination rate, and magnetic-field, using parameters we believe appropriate to the HgTe
quantum wells in Ref.~\cite{olshanetsky_metal_2013}. In the case of zero magnetic
field, we checked the validity of EMA by introducing a model of resistors on a
discrete lattice, which we solve numerically.

\textbf{Quantum well at equilibrium.---}We consider a 
two-dimensional electron and hole mixture with densities
$n_n(r)$ and  $n_p(r)$, respectively. The characteristic correlation
length scale of the disorder is set by the distance to charged 
impurities, which can typically be on the order of hundred nanometers.
We expect this scale to be much larger than both the microscopic mean-free
path of the charge carriers and the Coulomb screening length.
Therefore, we will take a model in which we treat Coulomb 
interactions to be of the local form
\begin{subequations}\begin{align}
 \phi_n(r) &= \int dr' K_c(r-r') n_n(r') = K n_n(r)  \\
 \phi_p(r) &= -\int dr' K_c(r-r') n_p(r') = -K n_p(r) \;,
\end{align}\label{eq:co}\end{subequations}
where $K_c(r-r')$ is the Coulomb kernel and $K$ is the effective Coulomb interaction parameter.
The precise length scales of the screened Coulomb interaction 
and disorder potential do not enter into our analysis.

We define the energies in our description according to the level 
scheme illustrated in \figc{fig:setting}{b}:
\begin{subequations}
\label{eq:muEq}
\begin{align}
 \mu_n^\text{ec}&=\tilde \mu_n(r) +\frac{\Delta}{2} + U(r) \\
 \mu_p^\text{ec}&=-\tilde \mu_p(r) -\frac{\Delta}{2} + U(r) \;,
\end{align}
\end{subequations}
where $\mu_n^\text{ec}$ ($\mu_p^\text{ec}$) is the electrochemical potential of the 
electrons (holes), $\tilde \mu_n(r)$ [$-\tilde \mu_p(r)$] is the 
electron [hole] chemical potential measured from the bottom of the conduction band 
[top of the valance band], and $\Delta$ is the distance between the edges
of the respective bands. Finite band overlap as indicated in \figc{fig:setting}{b}
corresponds to $\Delta <0$.
The electrical potential is defined as $U(r) = \phi_n(r) + \phi_p(r) + V(r)$, where $V(r)$ 
describes the smooth spatial randomness of the potential on scale $W$,
which gives rise to the puddle formation, \figc{fig:setting}{c}.

At equilibrium the electrochemical potentials of the electrons and 
holes are identical, \ie, $\mu_n^\text{ec}= \mu_h^\text{ec}= \mu^\text{ec}$ 
and determined by the gate voltage. Using \eqq{eq:co}{eq:muEq} we find
\begin{subequations}
\label{eq:cpsc}
\begin{align}
 \tilde \mu_n(r)&=\mu^\text{ec}-K[n_n(r) - n_p(r)] - V(r) - \frac{\Delta}{2} \\
 \tilde \mu_p(r)&=-\mu^\text{ec}+K[n_n(r) - n_p(r)] + V(r) -\frac{\Delta}{2} \;,
\end{align}
\end{subequations}
which have to be solved self-consistently in the presence of disorder as
$n_\alpha(r)$ itself depends on the chemical potential $\tilde\mu_\alpha(r)$.
\begin{figure}
\begin{center}
 \includegraphics[width=0.48\textwidth]{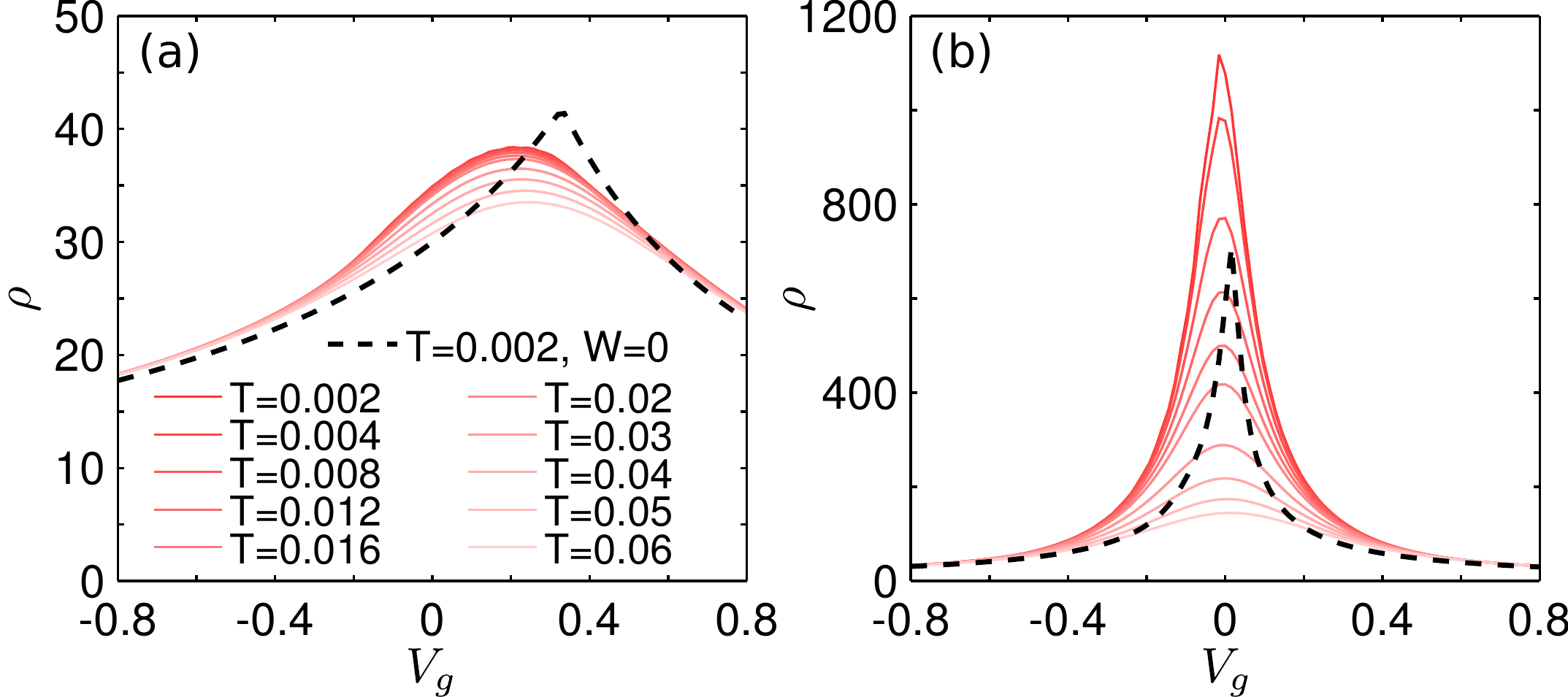}
\end{center}
\caption{\label{fig:r} (Color online) Resistivity $\rho$, \eq{eq:rho}, as a function of the gate 
voltage $V_g$ for different temperatures $T$, characteristic scale
of the potential fluctuations $W=1$, vanishing electron-hole recombination rate $\gamma=0$, 
and \fc{a} band overlap $\Delta=-0.5$ and \fc{b} $\Delta=-0.025$. 
Solid curves are evaluated for disordered systems using the effective medium approximation,
while the dashed line is evaluated for a clean system. For small band overlap 
\fc{b} the random local potential leads to an increase of resistivity.}
\end{figure}

\textbf{Non-equilibrium treatment.---}When a bias voltage is applied
to the electrodes at the edges of the sample, the local potential
$\delta U(r)$ changes in the entire sample thus causing a change of 
$\delta \phi_\alpha(r)$ and $\delta \mu^\text{ec}_\alpha(r)$.  Out of equilibrium the bulk 
electron/hole electrochemical potentials therefore differ from each other
\begin{subequations}
\begin{align}
 \delta \mu_n^\text{ec}(r)&=\delta \tilde \mu_n(r) +\delta U(r) \\
 \delta \mu_p^\text{ec}(r)&=-\delta \tilde \mu_p(r) +\delta U(r)\;.
\end{align}
\end{subequations}
At the boundary the electrochemical potentials of both components are identical and 
fixed by the potential imprinted from leads $\delta \mu^\text{ext}$.

The electron and hole currents are driven by the electrochemical potentials
\begin{align}
 \begin{pmatrix}   j_n \\ j_p  \end{pmatrix}= - \Sigma  \begin{pmatrix}  \nabla \delta \mu_n^\text{ec} \\ \nabla \delta \mu_p^\text{ec}  \end{pmatrix}   \;,                             
  \quad \Sigma = \begin{pmatrix}  \sigma_n & 0 \\ 0 &  \sigma_p\end{pmatrix}\;,
  \label{eq:jM}
\end{align}
where $\sigma_\alpha(r)$ is the microscopic conductivity whose functional form we 
derive in the supplemental material from a Boltzmann transport formalism~\cite{supp}.
In particular, we consider the linear current response of the two fluids to a small bias voltage. In that regime energy relaxation effects have not to be considered and thus we can assume that the system is locally at equilibrium. In our notation, $j_n$ and $j_p$ are vectors in the $x$-$y$ plane, while $\sigma_\alpha$ are scalars, in the absence of an applied magnetic field.

The non-equilibrium dynamics of the electron-hole channels is decoupled, \eq{eq:jM}. However, recombination processes 
of rate $\gamma$ dynamically couple the fluids, which can be taken into account by 
the continuity equation
\begin{equation}
 \frac{\partial \vec n}{\partial t} + \nabla \vec j = \Gamma \delta \vec \mu^\text{ec}\;,\quad \Gamma = \begin{pmatrix}  -\gamma & \phantom{-}\gamma \\ \phantom{-}\gamma &  -\gamma\end{pmatrix}\;.
\end{equation}
Here, we used a two-component vector notation with the electron and hole 
component at the first and second entry, respectively. The steady state is obtained from the continuity equation by setting $\partial \vec n/\partial t=0$
and boundary conditions that fix the electrochemical potential:
\begin{equation}
 \nabla \vec j - \Gamma \delta \vec \mu^\text{ec} = 0, \qquad \delta \mu_n^\text{ec}|_\text{bnd} = \delta \mu_p^\text{ec}|_\text{bnd}  = \delta \mu^\text{ext}\;.
 \label{eq:jMstst}
\end{equation}
Solving \eqw{eq:jMstst} amounts to determining the conductance of a random medium.
One approach is to discretize \eq{eq:jMstst} and map it onto a resistor network, see supplemental
material~\cite{supp}. Alternatively, one can exploit approximate features of such problems by a 
mean-field treatment, often referred to as EMA.

Cheianov, et al.~\cite{cheianov_random_2007} have employed a resistor network model, similar 
in some respects to the one introduced here, to derive critical exponents and scaling behavior 
of graphene near its neutrality point. They use a percolation network analysis, which if applied 
to the present model, could be used to obtain the singular behavior in the limit where $T$, $\Delta$, 
and $\gamma$ tend to zero. In contrast, EMA predictions for the critical behavior would be
qualitatively, but not quantitatively correct.

\textbf{Effective medium approximation.---}EMA considers inclusions, labeled 
with superscript $i$, that are embedded in an effective medium, labeled 
with superscript $m$.  The embedding is determined self-consistently by 
requiring that the current in the effective medium $\vec j^m = -\Sigma^m \langle \nabla \delta \vec \mu^\text{ec} \rangle$ is identical
to the average current in the sample $\langle \vec j^i \rangle = -\langle \Sigma^i \nabla\delta  \vec \mu^\text{ec} \rangle$,
for details see the supplemental material~\cite{supp}. 

The resistivity can be evaluated from the total current response of the system. Considering
that the electrochemical potential at the boundary is fixed \eqw{eq:jMstst}, we 
define the total resistivity $\rho$ as $j_n + j_p = \rho^{-1} \nabla  \delta \mu^\text{ext}$ with
\begin{equation}
 \rho^{-1} = \sum_{\alpha\beta} \Sigma^m_{\alpha \beta} \;,
 \label{eq:rho}
\end{equation}
where $\Sigma^m$ is the self-consistently determined conductivity matrix of the medium.

\textbf{Resistivity of disordered HgTe quantum wells.---}We now apply the method 
developed for a general disordered two-fluid model to HgTe quantum wells studied 
experimentally in~\cite{kvon_two-dimensional_2008,olshanetsky_scattering_2009,kvon_two-dimensional_2011,
olshanetsky_two-dimensional_2012,olshanetsky_metal_2013}.
In HgTe quantum wells a transition from a topological trivial insulator to a
quantum spin Hall insulator can be driven by enhancing the thickness of the 
well~\cite{bernevig_quantum_2006,konig_quantum_2007,hart_induced_2014}.
When further increasing the width of the well the system undergoes another 
transition to a semimetallic phase in which both electron and hole carriers 
contribute to transport~\cite{kvon_two-dimensional_2008,olshanetsky_scattering_2009,kvon_two-dimensional_2011,
olshanetsky_two-dimensional_2012,olshanetsky_metal_2013}. In the following, we consider 20nm HgTe quantum wells
grown in the (100)-direction as studied in Ref.~\cite{olshanetsky_metal_2013}. In that system, the effective electron and hole masses 
are very different $m_p/m_n \sim 6$~\cite{olshanetsky_two-dimensional_2012}, a unit cell contains one electron and
four hole pockets, and at atmospheric pressure the conduction and valance band overlap about $|\Delta|\sim 1.2$meV 
to $1.5$meV. In the experiment of Ref.~\cite{olshanetsky_metal_2013} hydrostatic 
pressure of $\sim15$kbar is applied to the sample which is expected to decrease the band overlap.

In accordance with these observations we choose the following parameters for our model: 
we set the Coulomb interaction parameter to $K=0.5 m_n^{-1}$, take into account the large difference 
in the effective electron and hole masses $m_p/m_n = 6$, and set $m_n=0.025$. We sample the local potential 
$V(r)$ from a uniform distribution of width $W=1$, which we use as unit of energy. 
The disorder strength $W$ is renormalized by the effective screening parameter $\tilde K=1+gK$, where
$g=(m_n+4m_p)/\pi$, yielding the effective disorder strength $\tilde W = W/\tilde K \sim 0.2$. Long-range disorder will influence the transport 
provided the band overlap $|\Delta|<\tilde W$. Based on these considerations, we choose two 
extreme limits for the band overlap $\Delta=-0.5$ and $\Delta=-0.025$ which should model
zero and high pressure in experiment~\cite{olshanetsky_metal_2013}.
The studied HgTe quantum well has an indirect bandstructure in which the extrema of the 
conduction and valance bands are at different wave vectors, see \figc{fig:setting}{b}. The electron-hole recombination 
would therefore require phonon scattering which is suppressed at low temperatures. Thus
we mostly consider zero carrier recombination rate $\gamma=0$.
\begin{figure}
\begin{center}
 \includegraphics[width=0.48\textwidth]{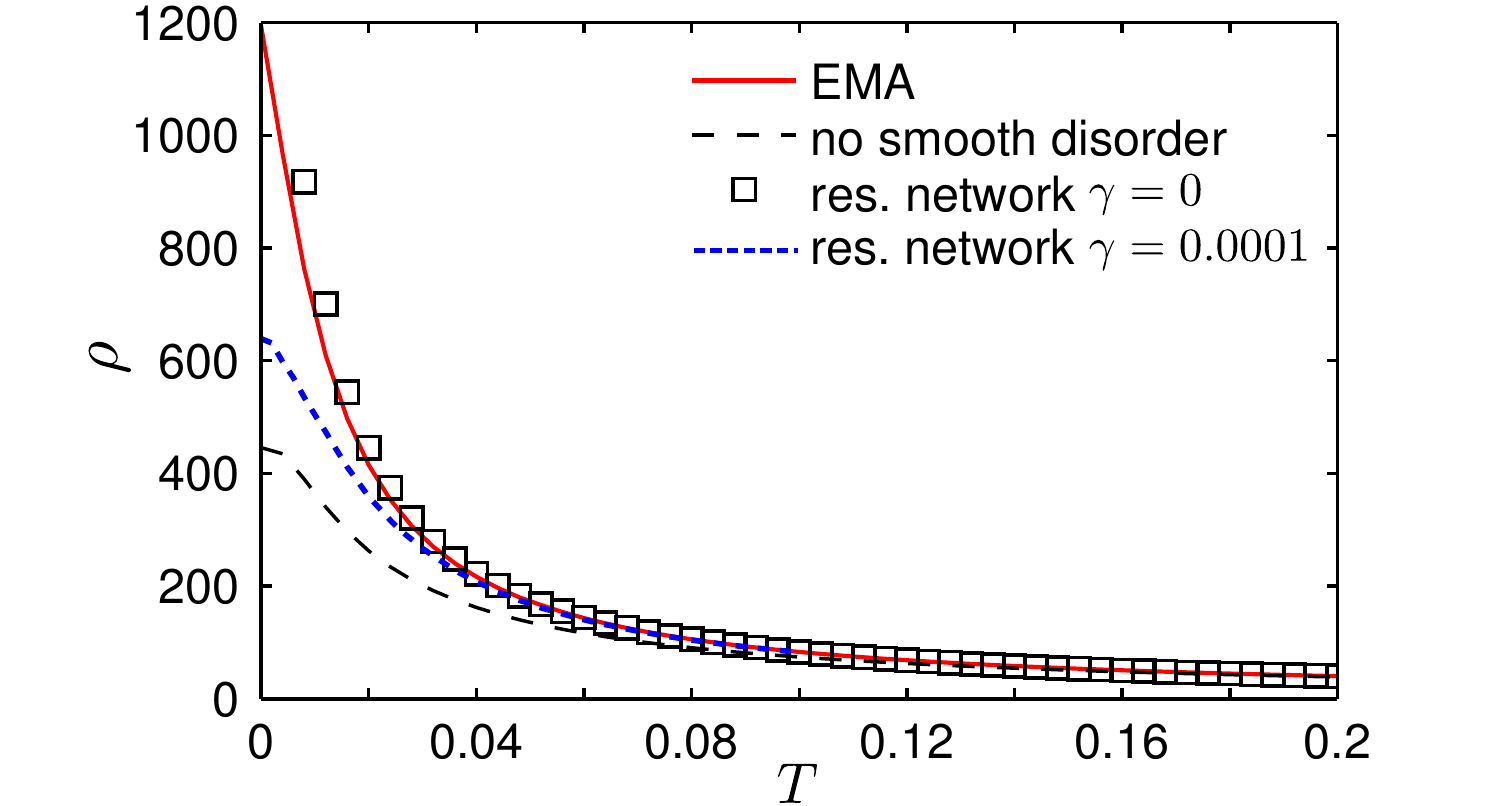}
\end{center}
\caption{\label{fig:rCmp} (Color online) Resistivity $\rho$ as a function of temperature $T$ for 
$V_g=-0.02$, $\Delta=-0.025$, $W=1$, and $\gamma=0$ evaluated with EMA, solid line, and resistor networks, squares. 
This data are compared with the resistivity $\rho$ in the presence of weak
electron hole recombination rate $\gamma=0.0001$, dotted line, and of the clean system $W=0$, dashed line. At low temperatures 
$\rho$ is considerably enhanced by disorder.}
\end{figure}

In \fig{fig:r} we show the resistivity obtained from EMA as a function 
of the gate voltage $V_g$ which directly modifies the electrochemical potential in \eq{eq:cpsc}. 
In accordance with the experiment~\cite{olshanetsky_metal_2013} we observe 
metallic behavior for $\Delta=-0.5$, with a weak dependence on temperature 
only \fc{a}. The asymmetry in the curves with respect to $V_g=0$ 
arises due to the large difference in electron and hole masses. For comparison we show the pure case $W=0$, dashed lines. 
For reduced band overlap $\Delta=-0.025$ and at $V_g\sim0$, the resistivity increases strongly at low temperatures 
while at high temperatures the system remains conducting \fc{b}. 
Further the maximum in the resistivity is shifted toward lower gate voltage. 
These results qualitatively explain several features of the HgTe quantum well experiments of 
Ref.~\cite{olshanetsky_metal_2013}.

The enhancement of the resistivity $\rho$ at low temperatures $T$  and the flattening 
out at high temperatures is demonstrated in \fig{fig:rCmp} for a fixed gate voltage $V_g$. 
In this plot we compare the results obtained within EMA to the solution of a resistor network~\cite{supp} of
size $L\times L=400\times400$ for vanishing electron-hole recombination rate 
$\gamma=0$ and find good agreement. Finite $\gamma > 0$ decreases the 
sharp low-temperature feature. We also compare the resistivity of the disordered systems 
to the resistivity of the clean system and find an enhancement at low temperatures, 
which is one of the main observations of our work. 

A percolation picture provides further insights.  When the recombination
rate is very small, we must compute separately the conserved currents of
electrons and holes, and add the results in the end.  Let us consider the
electron conductance as an example. If the band overlap is small, and the
system is electrically neutral, then the regions where only electrons
exist at $T=0$ will not percolate across the sample.  In order to get from
one of these regions to another, an electron has to cross the
intermediate region, where electrons and holes coexist, which will
generally occur at isolated junctions, where the two electron puddles
come close together, see \eg dashed rectangle in \figc{fig:setting}{c}. 
The conductance of these junctions will be small,
since they occur at places where the electron and hole densities are
both nearly vanishing.  The resistance of the electron network will be
dominated by these junctions, and in fact it will diverge in the limit
where the overlap goes to zero and there are equal numbers of electrons
and holes.  By contrast, at high temperatures, there will be a large
number of thermally excited electrons and holes even in the regions
separating the electron and hole dominated areas, so carriers can get across the
sample without crossing a region of low conductivity.  This physics is
correctly captured by EMA.

\textbf{Resistivity in the presence of magnetic field.---}To understand the relative electron 
and hole dominance we study the magnetotransport for fields applied perpendicular
to the transport layer. The model we derived for a setting with zero magnetic field is 
readily generalized to finite magnetic fields, see supplemental material~\cite{supp}
with the key difference that the resistivity has a tensorial structure consisting of a longitudinal $\rho_{xx}$ 
and a transverse $\rho_{xy}$ contribution~\cite{abrikosov_metals_1972,bergman_magnetotransport_1999}. 
\begin{figure}
\begin{center}
 \includegraphics[width=0.48\textwidth]{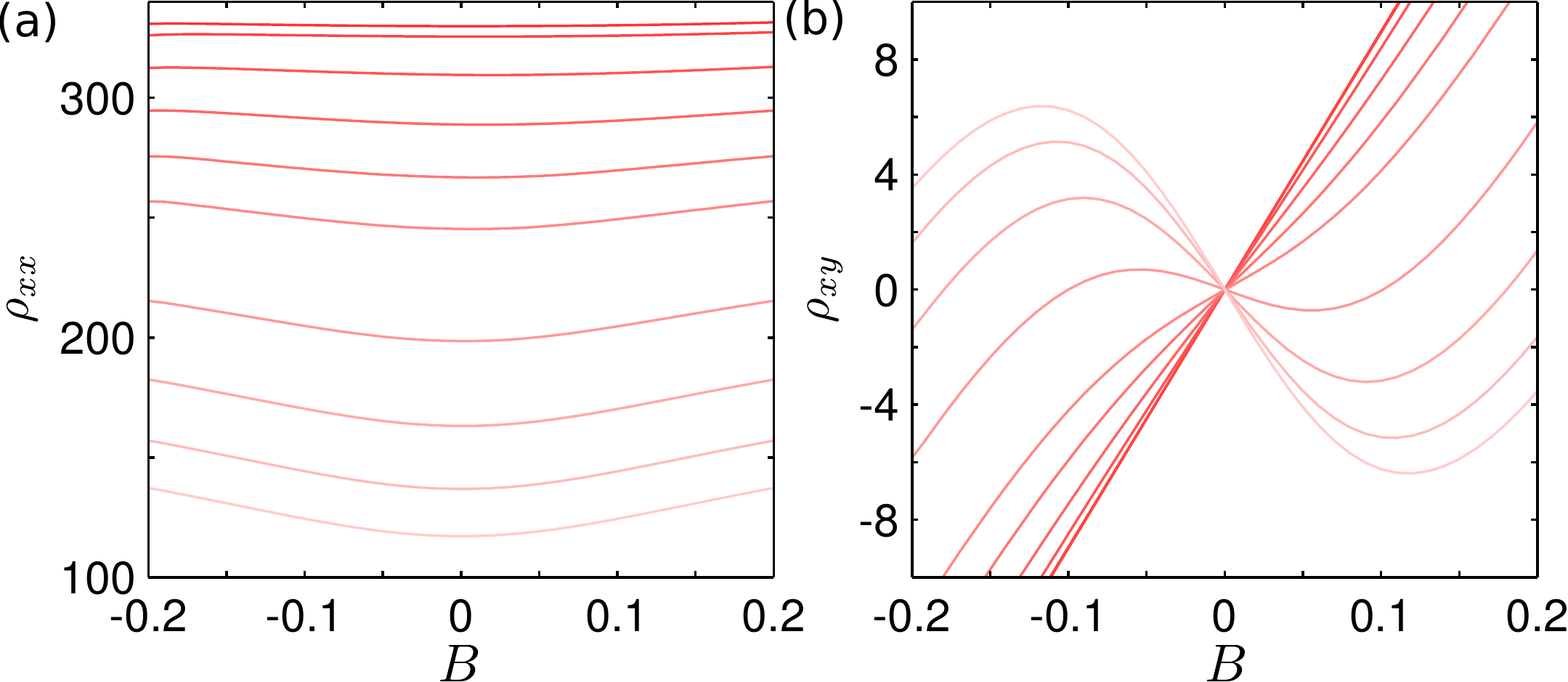}
\end{center}
\caption{\label{fig:rB} (Color online) \fc{a} Longitudinal $\rho_{xx}$ and \fc{b} transverse $\rho_{xy}$ resistivity
as a function of the magnetic field $B$ applied perpendicular to electron-hole mixture for the same temperatures $T$ 
as in \fig{fig:r}. The curves are taken at gate voltage $V_g=-0.16$, band overlap $\Delta=-0.025$, disorder strength $W=1$, 
and relaxation rate $\gamma=0$.}
\end{figure}

In \fig{fig:rB} we show the \fc{a} longitudinal $\rho_{xx}$ and the \fc{b} transverse $\rho_{xy}$ resistivity as 
a function of magnetic field $B$ for fixed gate voltage $V_g=-0.16$. We find that the 
longitudinal resistivity $\rho_{xx}$ increases with magnetic field $B$ and decreases with temperature $T$. 
For the chosen gate voltage the sign of the Hall charge (\ie, the slope of $\rho_{xy}$ at $B=0$) changes 
with temperature. The gate voltage is adjusted such that at low temperatures holes 
are the dominating charge carriers. With increasing temperature percolating paths open up faster
for the light electrons as compared to the heavy holes. Thus the electrons dominate transport at
high temperatures leading to the change of the Hall charge.

\textbf{Conclusions and outlook.---}We developed a theory for transport in long-range disordered 
two-fluid systems as realized for instance in semimetallic quantum wells, thin films, and bilayer
graphene. 

We applied the developed technique to study transport in HgTe quantum wells and found that it captures 
several characteristic features observed in experiment~\cite{olshanetsky_metal_2013}
including the strong enhancement of the resistivity at low temperatures near charge neutrality. 
In Ref.~\cite{olshanetsky_metal_2013} the authors proposed an alternative explanation 
of this effect based on the phase transition to an excitonic insulator at low temperatures
which does not consider long-range disorder but rather requires strong interactions. 
In contrast, in our theory which identifies long-range disorder as a crucial mechanism, the sharp 
enhancement of the resistivity is not indicative of a true phase transition. 
At low temperatures the resistivity will rather saturate, albeit at a very large value. 
This feature is generic for semimetals with small band overlap and results from
(i) the relatively small percolating portion of coexisting electron and hole states and (ii) the
vanishing electron-hole recombination rate relevant to the indirect bandstructures, such as
the one of the considered HgTe quantum well.

This leads us to the conclusion that large length-scale disorder is a central mechanism in 
these experiments. Of course a complex interplay between long-range disorder and
interactions is conceivable as well. Further experimental studies are therefore needed to fully confirm 
the picture. In particular, it would be interesting to explore the resistivity as a function 
of the applied pressure, which should tune the band overlap continuously, and thus allow
to study the emergence of the strong enhancement of the resistivity at low temperatures.

\textbf {Acknowledgements.---}We thank D. Kvon for introducing us to this
problem and for sharing experimental results prior to publication.
The authors acknowledge support from Harvard-MIT CUA, 
ARO-MURI Quism program, ARO-MURI on Atomtronics, STC Center for Integrated 
Quantum Materials, NSF grant DMR-1231319,
as well as the Austrian Science Fund (FWF) Project No. J 3361-N20.

\end{document}


\title{Supplemental Material for\\Transport in two-dimensional disordered semimetals}

\author{Michael Knap}
\email[]{knap@physics.harvard.edu}
\affiliation{Department of Physics, Harvard University, Cambridge MA 02138, USA}
\affiliation{ITAMP, Harvard-Smithsonian Center for Astrophysics, Cambridge, MA 02138, USA}

\author{Jay D. Sau}
\affiliation{Department of Physics, Harvard University, Cambridge MA 02138, USA}
\affiliation{Department of Physics, Condensed Matter Theory Center and the Joint Quantum Institute, University of Maryland, College Park, MD 20742, USA}

\author{Bertrand I. Halperin}
\affiliation{Department of Physics, Harvard University, Cambridge MA 02138, USA}

\author{Eugene Demler}
\affiliation{Department of Physics, Harvard University, Cambridge MA 02138, USA}

\date{\today}

\begin{abstract}

\end{abstract}


\maketitle

\onecolumngrid

\section{Transport coefficients within Boltzmann formalism}

Using the Boltzmann formalism we calculate the functional forms of the density $n_\alpha(r)$, the
microscopic conductivity $\sigma_\alpha(r)$, the diffusion constant $D_\alpha(r)$, and the compressibility $\kappa_\alpha(r)$. In 
two-dimensions these quantities can be evaluated in closed forms as the density of states is constant. For the electron
and hole densities we find
\begin{subequations}\begin{align}
 n_n(r) &= \frac{m_n}{\pi} T \log (1+e^{\tilde \mu_n(r)/T}) \\
 n_p(r) &= 4\frac{m_p}{\pi} T \log (1+e^{\tilde \mu_p(r)/T}) \;,
\end{align}\label{eq:n}\end{subequations}
where $m_\alpha$ is the mass and $T$ the temperature. The chemical potentials $\tilde \mu_\alpha(r)$ depend implicitly on the
densities, see \eqt{3}. In case of HgTe quantum wells grown in $(100)$ direction the bandstructure consists of four 
hole pockets, which surround a single electron pocket~\cite{olshanetsky_metal_2013}, and thus the particle density $n_p$ is multiplied 
by a factor four. The densities depend parametrically on the the temperature $T$ and on the gate voltage $V_g$, which 
enters through the electrochemical potential.

For the conductivity $\sigma_\alpha$ we obtain 
\begin{align}
 \sigma_\alpha &= e \nu_\alpha n_\alpha\;,
\end{align}
where $\nu_\alpha$ is the mobility of the carrier and $e$ the charge of the electron.

\section{Effective Medium Approximation for the two-fluid model}
\begin{figure}[b]
\begin{center}
 \includegraphics[width=0.3\textwidth]{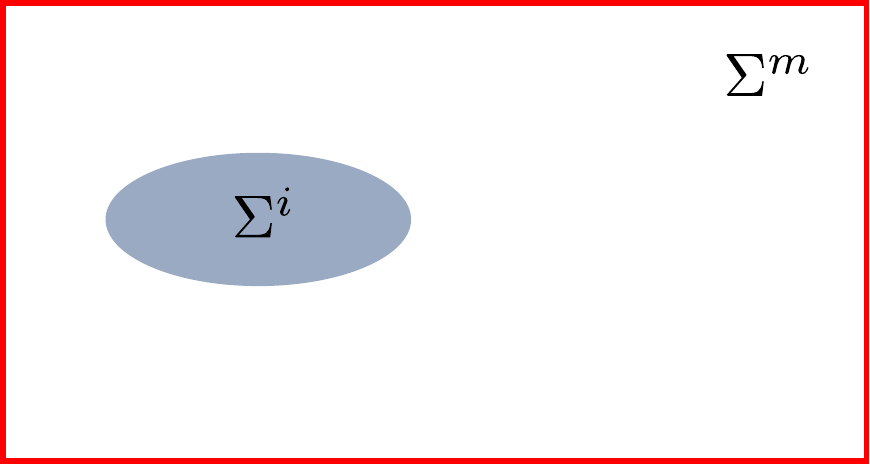}
\end{center}
\caption{\label{fig:ema} The effective medium approximation self-consistently
embeds an inclusion with microscopic conductivity matrix $\Sigma^i$ into a medium characterized by $\Sigma^m$.}
\end{figure}

\textbf{Self-consistency condition.---}EMA self-consistently embeds an inclusion, labeled $i$, in an effective medium, labeled $m$, see 
\fig{fig:ema}. We impose the self-consistency condition that the 
current of the effective medium is identical to the average current in the sample
\begin{equation}
 \langle {j}^i \rangle = {j}^m \;.
 \label{eq:sc}
\end{equation}
We introduce the microscopic conductivity of the inclusion $\Sigma^i=\Sigma^m+\delta \Sigma^i$ and the
effective medium $\Sigma^m$, respectively.
Thus $\Sigma(r)$ as a function of space is given by
\[
 \Sigma(r) = \Sigma^m+\Theta^i(r) \delta \Sigma^i(r) \;,
\]
where $\Theta^i(r)=1$ if $r$ is in the inclusion and $\Theta^i(r)=0$ otherwise. 
From the continuity equation
\begin{equation}
  \nabla \vec j = -\nabla (\Sigma \nabla \delta \vec \mu^\text{ec})=\Gamma  \delta \vec \mu^\text{ec} \;.
 \label{eq:stst}
\end{equation}
we find
\begin{equation}
 \Sigma^m_{\alpha\beta} \nabla^2 \delta \mu^\text{ec}_\beta(r) + \nabla[\Theta^i(r) \delta \Sigma_{\alpha\beta}^i(r) \nabla \delta \mu^\text{ec}_\beta(r)] + \Gamma_{\alpha \beta} \delta \mu^\text{ec}_\beta(r) = 0 \;.
 \label{eq:de}
\end{equation}
We define the Green's function for the homogeneous system
\begin{equation}
 (\Sigma^m_{\alpha\beta} \nabla^2 + \Gamma_{\alpha \beta}) G_{\beta\gamma}(r-r') + \delta(r-r')\delta_{\alpha\gamma} = 0 \;.
 \label{eq:gf}
\end{equation}
With that we can write the formal solution of the differential equation \eqw{eq:de}
\[
 \delta \mu^\text{ec}_\alpha = \delta \mu_\alpha^\text{ext} + \int d^2 r' G_{\alpha \beta}(r-r') \nabla' (\Theta^i \delta \Sigma_{\beta \gamma}^i \nabla' \delta \mu^\text{ec}_\gamma) \;.
\]
Integrating by parts and taking the spatial derivative gives
\begin{align}
-\nabla \delta \mu^\text{ec}_\alpha &= -\nabla \delta \mu_\alpha^\text{ext} 
+ \int_{r' \in i} d^2 r' \nabla\nabla'  G_{\alpha \beta}(r-r')  \delta \Sigma_{\beta \gamma}^i \nabla' \delta \mu^\text{ec}_\gamma  \;.
\label{eq:emascCond}
\end{align}

In the following we will first discuss the solution of the static problem in the limit of
vanishing electron-hole recombination rate $\gamma=0$ and after that consider the problem
with finite recombination rate $\gamma\neq0$.

\textit{Case of $\gamma=0$.---}In the limit of zero electron-hole recombination 
rate $\gamma=0$ the field in the inclusion $\nabla \delta \mu^\text{ec}_\alpha = \text{const}$.
and the integral in \eq{eq:emascCond} can be evaluated exactly
\[
 \int_{r' \in i} d^2 r' \nabla\nabla'  G_{\alpha \beta}(r-r')  = \frac{1}{2} (\Sigma^m)^{-1}_{\alpha \beta} \;
\]
yielding
\begin{equation}
 [\mathds{1} + \frac{1}{2} (\Sigma^m)^{-1} \delta \Sigma^i ]\nabla \delta \vec \mu^\text{ec} = \nabla \delta  \vec \mu^\text{ext}\;.
 \label{eq:gam0}
\end{equation}
Thus for elliptic inclusions and $\gamma=0$ the external field $\nabla \delta \vec \mu^\text{ext}$ generated from the potential difference 
in the leads is related to the electrochemical potential in the inclusion through
\begin{equation}
\nabla \delta \vec \mu^\text{ec}_\alpha 
= \Lambda^i_{\alpha\beta}\nabla \delta \vec \mu^{\text{ext}}_\beta\;
 \label{eq:lambda}
\end{equation} 
with $\Lambda^i$ given by \eq{eq:gam0}. Using this relation we can express the currents as
\begin{subequations}
\begin{align}
 \vec j^i &=  -\Sigma^i \Lambda^i \nabla \delta \vec \mu^\text{ext} \label{eq:ji}\\
 \vec j^m &= -\Sigma^m \langle  \Lambda^i\rangle \nabla \delta \vec \mu^\text{ext}\label{eq:jm}\;,
\end{align}
\end{subequations}
which yields the EMA self-consistency condition 
\begin{align}
  &\Sigma^m  \langle \Lambda^i\rangle= \langle\Sigma^i \Lambda^i\rangle \label{eq:emasc} 
\end{align}
for the two-fluid model.

\textit{Case of $\gamma\neq0$.---}We consider finite electron-hole recombination rate $\gamma$. 
Using in \eq{eq:emascCond} the definition of the Green's function \eqw{eq:gf} we find
\[
 -\nabla\delta \vec \mu^\text{ec} = -\nabla\delta \vec \mu^\text{ext} + \frac{1}{2} (\Sigma^m)^{-1} \left\lbrace\delta\Sigma^i \nabla \delta \vec \mu^\text{ec}+
 \Gamma \int d^2 r' G(r-r')  \delta\Sigma^i \nabla' \delta \vec \mu^\text{ec}\right\rbrace \;.
\]
The solution of the Green's function \eq{eq:gf} can be obtained by the following manipulations
\begin{align*}
 G(k) &= (\Sigma^m k^2-\Gamma)^{-1} = (\Gamma^{-1} \Sigma^m k^2-\mathds{1})^{-1}\Gamma^{-1} \\
 &=-U(E k^2+\mathds{1})^{-1}U^{-1}\Gamma^{-1} = -U \tilde G (k) U^{-1}\Gamma^{-1}\;.
\end{align*}
In the first line, we transformed to Fourier space, while in the second
line we introduced the unitary transformation $-\Gamma^{-1} \Sigma^m U = U E$, which diagonalizes
$\Gamma^{-1} \Sigma^m$, and 
defined the Green's function $(-E \nabla^2+\mathds{1})\tilde G (r-r') = \mathds{1}\delta(r-r')$.
This is the Green's function of a screened Poisson equation (or similarly Helmholtz equation) which in two-dimensions
is of the form 
\[
G(r-r')=-\frac{1}{2\pi}E^{-1} K_0(\frac{|r-r'|}{\sqrt{E}})\;
\]
where $K_0(x)$ is the modified Bessel function of the second kind. With that we can write the formal 
solution relating $\nabla\delta \vec \mu^\text{ec}$ with $\nabla\delta \vec \mu^\text{ext}$
\begin{equation}
 -\nabla\delta \vec \mu^\text{ec} = -\nabla\delta \vec \mu^\text{ext} + \frac{1}{2} (\Sigma^m)^{-1} 
 \left\lbrace\delta\Sigma^i \nabla \delta \vec \mu^\text{ec}+\Gamma U \frac{E^{-1}}{2\pi}\int d^2 r' K_0\left(\frac{|r-r'|}{\sqrt{E}}\right) U^{-1} \Gamma^{-1} \delta\Sigma^i \nabla' \delta \vec \mu^\text{ec}\right\rbrace \;.
\end{equation}
This equation has to be solved self consistently. Well outside the inclusion $r\gg R$, the 
contribution from the integral decays to zero as the asymptotic limit of the modified Bessel function 
is $\sim\exp(-r)/\sqrt{r}$, which is consistent with the $\gamma=0$ results. 
We can make further analytical progress by taking the limit of small electron-hole recombination rate, which is 
of our primary interest. To this end, we perturbatively expand around the homogeneous $\gamma=0$ solution \eqw{eq:gam0}
\begin{align*}
 -\nabla\delta \vec \mu^{\text{ec},(1)} = -\nabla\delta \vec \mu^\text{ext} + \frac{1}{2} (\Sigma^m)^{-1} 
 \bigg\lbrace\mathds{1}+\Gamma U \underbrace{\frac{E^{-1}}{2\pi}\int_R d^2 r' 
 K_0\left(\frac{|r-r'|}{\sqrt{E}}\right)}_{= M(r|R,\gamma)} U^{-1} \Gamma^{-1} \bigg\rbrace \delta\Sigma^i \nabla \delta \vec \mu^{\text{ec},(0)}\;.
\end{align*}
\begin{figure}
\begin{center}
 \includegraphics[width=0.7\textwidth]{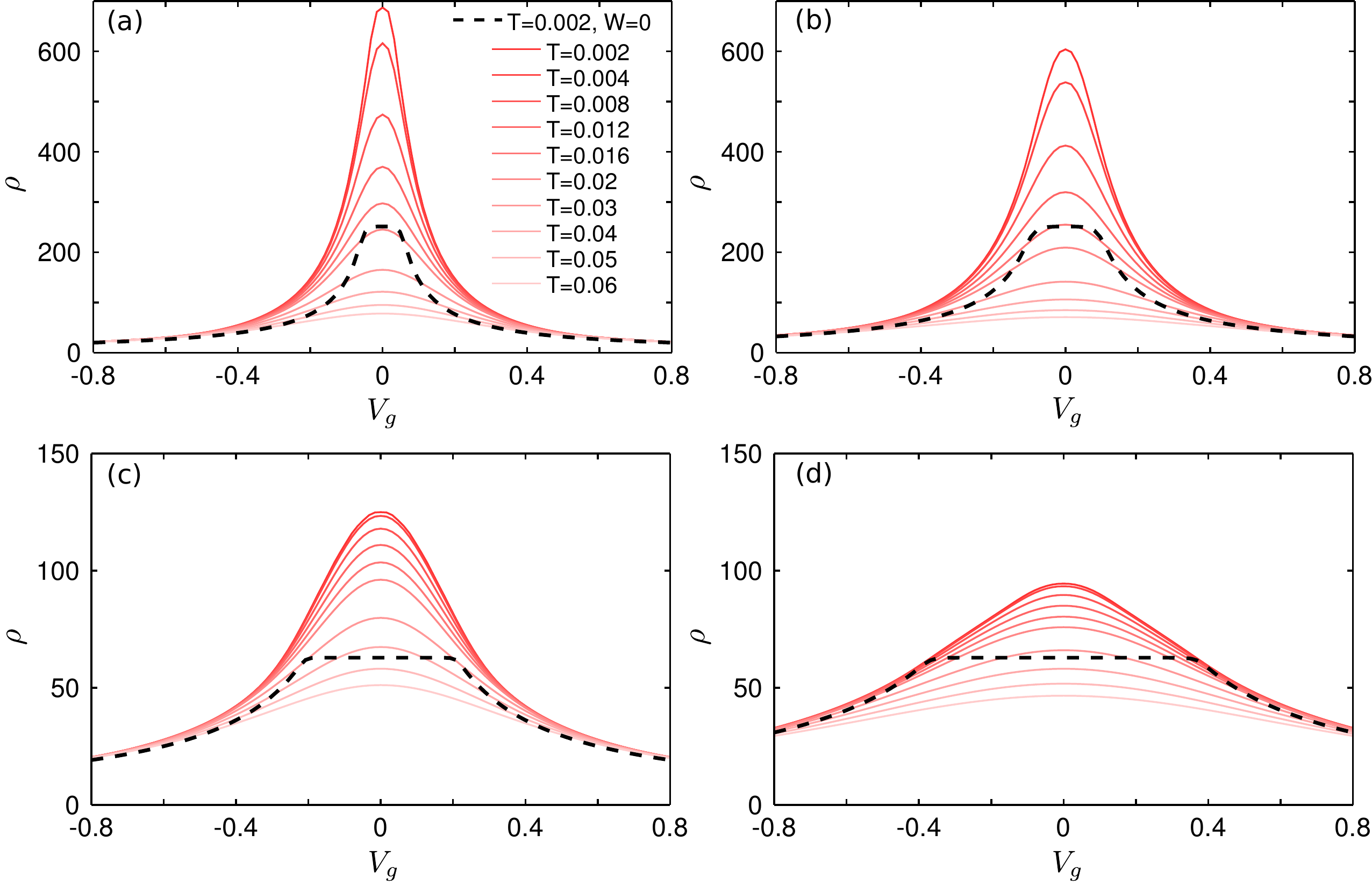}
\end{center}
\caption{\label{fig:rSupp} Resistivity $\rho$ as a function of the gate voltage $V_g$ for the electron-hole symmetric
case ($m_n=m_p=0.5$, same number of electron and hole pockets), $W=1$, $\gamma=0$, and \fc{a} $K=5m_n^{-1}$, $\Delta=-0.025$, \fc{b} $K=10m_n^{-1}$, 
$\Delta=-0.025$, \fc{c} $K=5m_n^{-1}$, $\Delta=-0.1$, and \fc{d} $K=10m_n^{-1}$, $\Delta=-0.1$. The resistivity of the clean system for
the lowest temperature is indicated by the dashed line. }
\end{figure}
\begin{figure}
\begin{center}
 \includegraphics[width=0.7\textwidth]{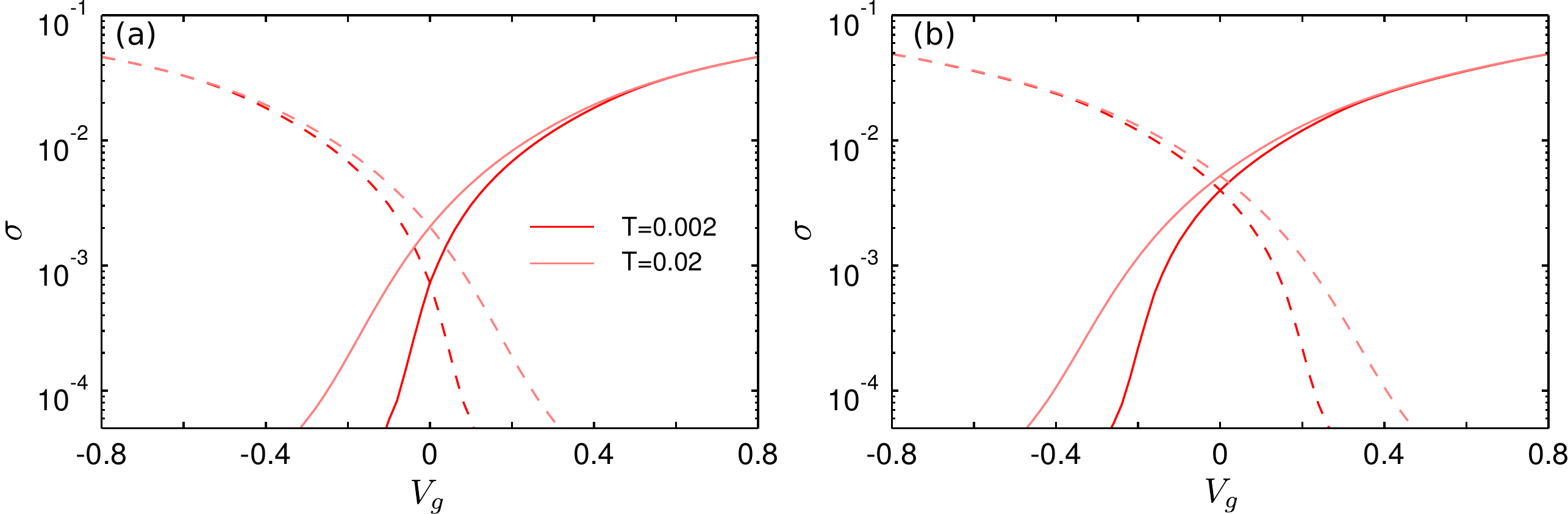}
\end{center}
\caption{\label{fig:sSupp} Electron conductivity, solid line, and hole conductivity, dashed line, as a function of the gate voltage $V_g$ for decoupled fluids, $\gamma=0$, at the electron-hole symmetric point ($m_n=m_p=0.5$, same number of electron and hole pockets), $W=1$, $K=5m_n^{-1}$, and \fc{a} $\Delta=-0.025$ and \fc{b} $\Delta=-0.1$ for two different temperatures. At large negative gate voltage holes dominate the transport while at large positive gate voltage electrons dominate. }
\end{figure}

Plugging in the zeroth order solution \eqw{eq:gam0} gives to first order in $\gamma$
\begin{align}
 \nabla\delta \vec \mu^{\text{ec},(1)} = \left\lbrace \mathds{1} - \frac{1}{2} (\Sigma^m)^{-1} 
 \big[\mathds{1}+\Gamma U  M(r|R,\gamma) U^{-1} \Gamma^{-1} \big] \delta\Sigma^i \big[\mathds{1} + 
 \frac{1}{2} (\Sigma^m)^{-1} \delta \Sigma^i \big]^{-1} \right\rbrace \nabla\delta \vec \mu^\text{ext}\;.
 \label{eq:lambdaGam}
\end{align}
In that case find again that $\nabla \delta \vec \mu^\text{ec}= \Lambda^i\nabla \delta \vec \mu^{\text{ext}}$ and thus
the EMA self-consistency condition \eqw{eq:emasc} as in the case of $\gamma=0$ but with a modified 
$\Lambda^i$ given by \eq{eq:lambdaGam}. In this work we evaluate the EMA equations only for vanishing 
electron-hole recombination rate $\gamma=0$. When solving the EMA equations at finite $\gamma\neq 0$, which we 
derived in this paragraph, 
the solution depends on the radius of the inclusions $R$ over which the average has to be performed
in addition to the average over the disorder distribution. 

\textbf{Supplemental results.---}Additional results for the electron-hole symmetric setting ($m_n=m_p=0.5$, and 
identical number of electron and hole pockets) and $\gamma=0$ obtained with EMA are shown in \figg{fig:rSupp}{fig:sSupp}.

\section{Effective Medium Approximation for the two-fluid model with magnetic field}

\textbf{Self-consistency condition.---}The current of charged particles subject to
an electric and a magnetic field can be derived within Boltzmann formalism~\cite{abrikosov_metals_1972}. 
For weak magnetic fields $\nu_\alpha |\vec B| \ll 1$ we find for the particle and hole 
current
\begin{subequations}
\begin{align}
 \vec j_n - \nu_n \vec j_n \times \vec B &= -\sigma_n  \nabla \delta \mu^\text{ec}_n  \\
 \vec j_p +\nu_p \vec j_p \times \vec B &= -\sigma_p \nabla \delta \mu^\text{ec}_p \;.
\end{align}
\end{subequations}
For zero magnetic field these equations reduce to the ones discussed in the previous section. 
\begin{figure}
\begin{center}
 \includegraphics[width=0.7\textwidth]{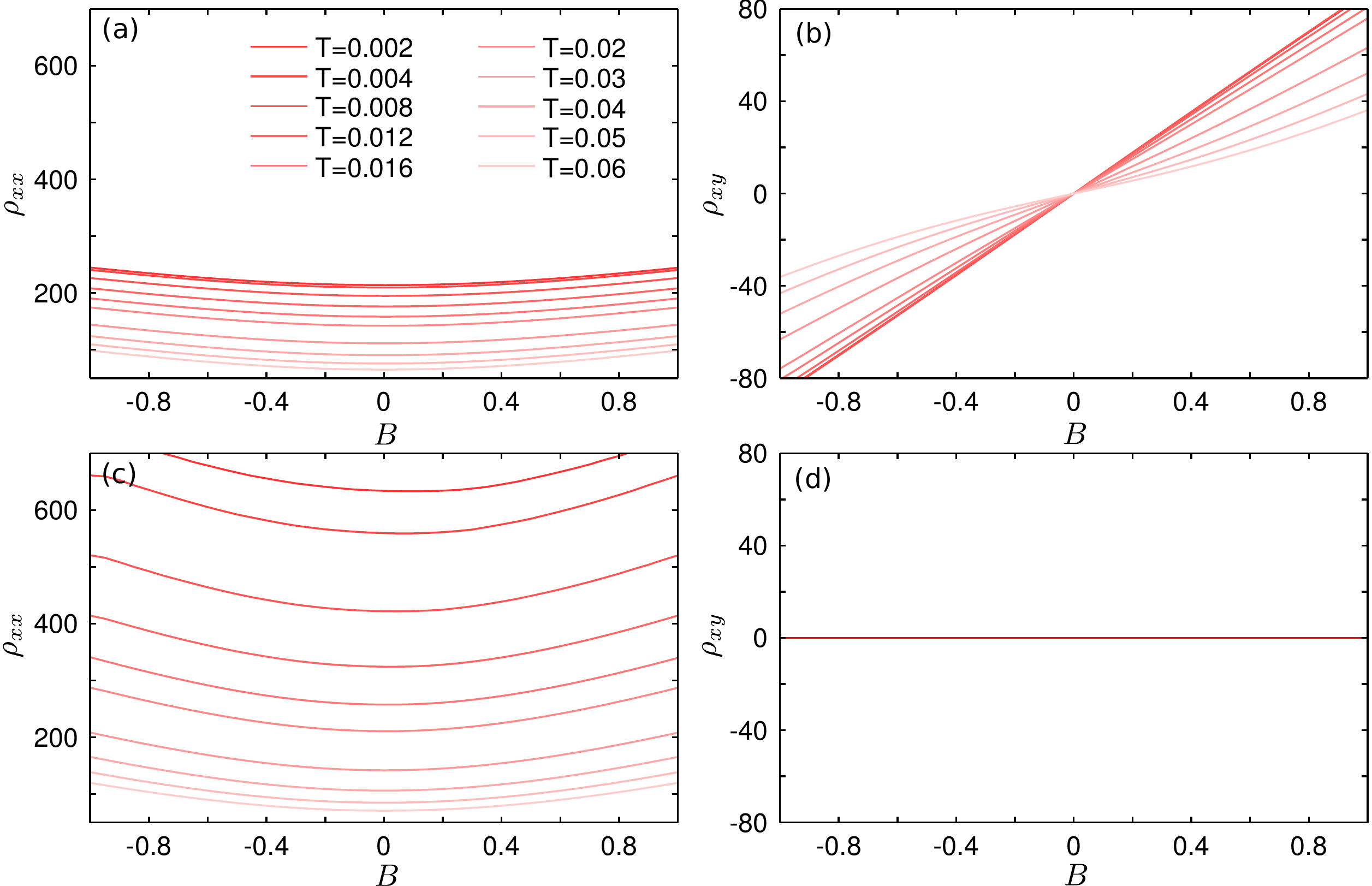}
\end{center}
\caption{\label{fig:rBSupp} Longitudinal $\rho_{xx}$ and transverse $\rho_{xy}$ resistivity as a 
function of magnetic field $B$ and temperature $T$ for the electron-hole symmetric
case ($m_n=m_p=0.5$, same number of electron and hole pockets), $K=10m_n^{-1}$, $\Delta=-0.025$, $W=1$, $\gamma=0$ and \fc{a,b} $V_g=-0.2$, \fc{c,d} $V_g=0$.  }
\end{figure}

When the magnetic field points perpendicular to the transport layer, these equations simplify and
we define the effective microscopic conductivity matrix $\tilde \Sigma$ as~\cite{abrikosov_metals_1972,bergman_magnetotransport_1999}
\begin{equation}
 \vec j = - H^{-1} (\Sigma \otimes \mathds{1}) \nabla \delta \vec \mu^\text{ec} = - \tilde \Sigma  \nabla \delta \vec \mu^\text{ec}\;.
 \label{eq:jB}
\end{equation}
where we used a four component notation with electron $x$, $y$ components as first an second entry 
and the hole $x$, $y$ components as third and fourth entry and $\mathds{1}$ is the $2\times2$
identity matrix in $x,y$ space. The matrix $H$ is
\begin{equation}
 H = \begin{pmatrix}
      H_n & 0 \\ 0 & H_p
     \end{pmatrix}, 
     \label{eq:H}
\end{equation}
with 
\[
 H_n = \begin{pmatrix}
      1 & -\nu_nB \\ \nu_n B & 1
     \end{pmatrix} \quad \text{and} \quad  H_p = \begin{pmatrix}
      1 & \nu_pB \\ -\nu_p B & 1
     \end{pmatrix}\;.
\]

Similarly as in the case of zero magnetic field, an inclusion with conductivity $\tilde \Sigma^i$ 
has to be embedded self-consistently in an effective medium with conductivity $\tilde \Sigma^m$. From that
the longitudinal $\rho_{xx}$ and transverse $\rho_{xy}$ resistivity can be obtained by tracing out
the two carrier types
\begin{equation}
 \begin{pmatrix}  \rho_{xx} & \rho_{xy} \\  -\rho_{xy} & \rho_{xx} \end{pmatrix}^{-1} = \sum_{\alpha\beta} \tilde \Sigma^m_{\alpha\beta} \;. 
 \label{eq:rhoB}
\end{equation}

At the steady state we have the differential equation
\begin{equation}
 \nabla\tilde \Sigma^m \nabla \delta \vec\mu^\text{ec} + \nabla[\Theta^i(r) \delta \tilde \Sigma^i(r) \nabla \delta \vec\mu^\text{ec}(r)] + (\Gamma\otimes\mathds{1}) \delta \vec \mu^\text{ec} = 0 \;.
 \label{eq:deB}
\end{equation}
The definition of the Green's function is thus 
\[
 (\nabla \tilde \Sigma^m \nabla + \Gamma\otimes\mathds{1})G(r-r') + \delta(r-r') \mathds{1}= 0 \;.
\]
However, here, only the part of the conductivity matrix $\tilde \Sigma$ which 
is symmetric in real space is relevant~\cite{bergman_magnetotransport_1999}. To determine this part we first express
\[
 \tilde \Sigma = \begin{pmatrix}
                  H_n^{-1}  \sigma_n & 0 \\
                  0 &  H_p^{-1}   \sigma_p \\
                 \end{pmatrix}\;,
\]
where
\[
 H_n^{-1} =  h_n^{-1} \begin{pmatrix}
                                     1 & \nu_n B \\ - \nu_n B & 1
                                    \end{pmatrix}\;,\quad H_p^{-1} =  h_p^{-1} \begin{pmatrix}
                                     1 & -\nu_p B \\  \nu_p B & 1
                                    \end{pmatrix}
\]
and $h_\alpha=1+\nu_\alpha^2 B^2$. Therefore, the symmetric contribution reduces to
\begin{equation}
 \tilde \Sigma^s = \Sigma^s  \otimes \mathds{1} = \begin{pmatrix}
                    { \sigma_n}/{h_n} &  0 \\
                    0 &  { \sigma_p}/{h_p}
                   \end{pmatrix} \otimes \mathds{1}\;.
\end{equation}
Since $ \tilde \Sigma^s$ does not depend on space, we have for the  Green's function 
\begin{equation}
  (\tilde \Sigma^{s,m} \nabla^2+ \Gamma\otimes\mathds{1}) G(r-r') + \delta(r-r') \mathds{1}= 0 \;.
\end{equation}

The differential equation \eqw{eq:deB} can now be evaluated in the same way as for
zero magnetic field. In particular, we find in the limit $\gamma=0$
\begin{equation}
 (\Lambda^i)^{-1} = \mathds{1} + \frac{1}{2} [(\Sigma^{s,m})^{-1}\otimes \mathds{1}](\tilde \Sigma^i - \tilde \Sigma^m)\;.
 \label{eq:gB}
\end{equation}

\textbf{Supplemental results.---}Additional results of the longitudinal and transverse resistivity 
for the electron-hole symmetric setting ($m_n=m_p=0.5$, and identical number of electron and hole 
pockets) and $\gamma=0$ obtained with EMA are shown in \fig{fig:rBSupp}.

\section{Resistor network}

Complimentary to solving the problem with EMA, we calculate the exact solution
of the transport problem at zero magnetic field, $B=0$, by mapping it onto a resistor network. On the bonds $b$ between two 
neighboring sites of this network, we introduce conductivity matrices 
$\Sigma(x, b)$ ($x$ is an arbitrary node of the resistor network). $\Sigma(x, b)$ is 
defined as in \eqt{5} and characterized by a local potential $V(x,b)$
drawn from a random distribution. Correlations between electrons and holes arise since
both constituents exhibit the same local random potential. 

In the bulk, the continuity equation $\nabla \vec j=\Gamma \delta \vec \mu^\text{ec}$ holds. Discretizing this 
condition gives
\begin{align}
 \forall x: \quad \vec 0&=\sum_{b} \vec{j}( x, b) - \Gamma \delta \vec \mu^\text{ec}(x) \nonumber\\
 &=  \sum_{b} \Sigma(x, b) [ \delta \vec\mu^\text{ec}({x+b})- \delta \vec\mu^\text{ec}({x})] + \Gamma \delta \vec \mu^\text{ec}(x)\;,
\end{align}
where in the second line we used that $\vec j = -\Sigma \nabla\delta  \vec \mu^\text{ec}$.
On the left and the right boundary, the electrochemical potential has a finite value
which imprints the bias voltage onto the system yielding the boundary conditions
\begin{align}
 \pm\delta \vec \mu &=\begin{pmatrix}
              1  & 0 \\ 0 & 1
             \end{pmatrix} \delta \vec \mu^\text{ec}(x) \;,	
 \label{eq:rnbnd}
\end{align}
where $x$ lies on the left,$+$ (right,$-$) boundary. Combining all the equations above we find for a 
resistor network on a two-dimensional hypercubic lattice of size $L\times L$, $2L^2$ linear equations. The system of equations
is mostly homogeneous, with the only exception of $4L$ boundary terms.
\begin{figure}
\begin{center}
 \includegraphics[width=0.4\textwidth]{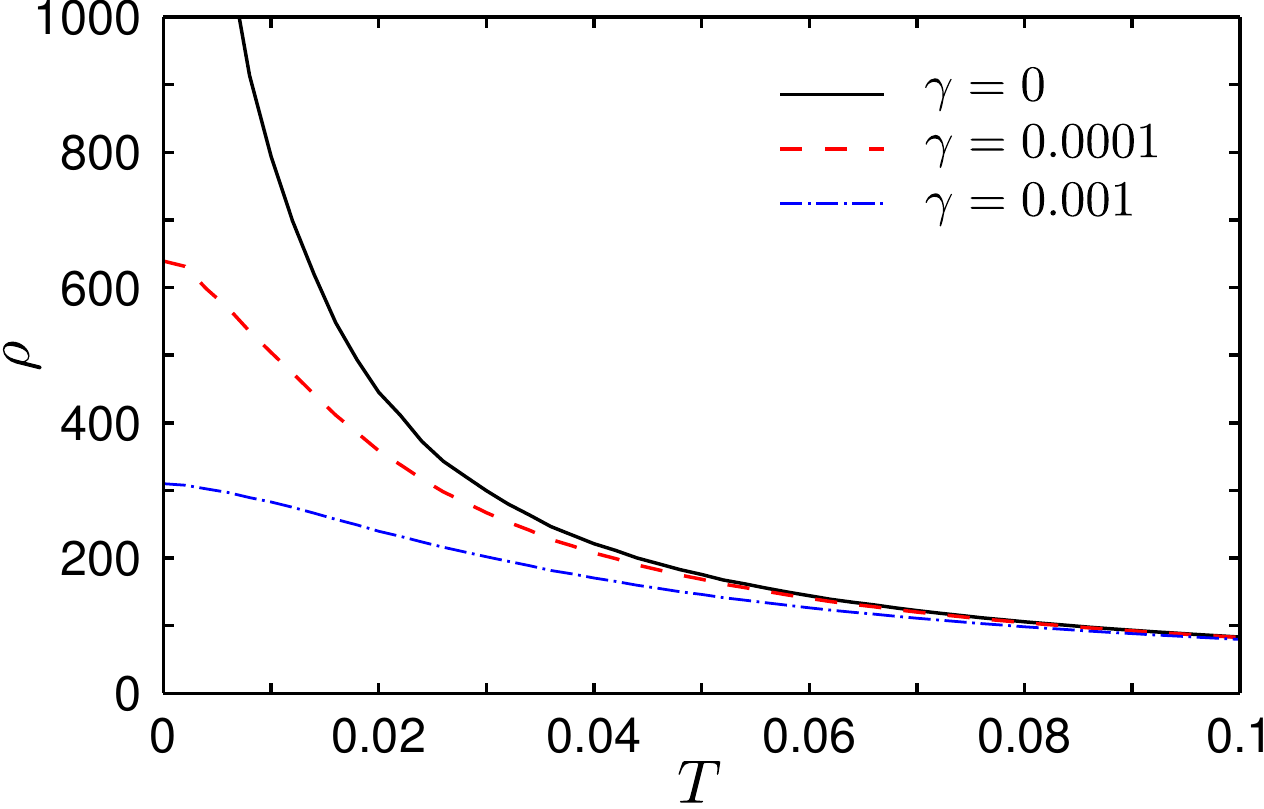}
\end{center}
\caption{\label{fig:resNet} Resistivity $\rho$ as a function of temperature $T$ evaluated 
from a resistor network of size $L\times L=400 \times 400$ for the same parameters as in \figt{3}, but
for different values of the electron-hole recombination rate $\gamma$. The recombination of electrons and holes 
leads to a suppression of the resistivity at low temperatures.}
\end{figure}

In \fig{fig:resNet} we show the resistivity for different values of the electron-hole recombination rate $\gamma$ and otherwise 
identical parameters as in \figt{3}. The result of the electron-hole recombination is a suppression of 
the resistivity at low temperatures. A comparison between EMA and the resistor network is included in \figt{3} and 
discussed in the main text.